\documentclass[12pt,preprint]{aastex}

\newcommand{\mbar}{\bar m}

\begin{document}
\title{Cosmological Acceleration Through Transition\\
to Constant Scalar Curvature}
\author{Leonard Parker\altaffilmark{1},
William Komp\altaffilmark{2},
and Daniel A. T. Vanzella\altaffilmark{3}}
\affil{Physics Department, University of
Wisconsin-Milwaukee,\\ Milwaukee, WI 53201, USA
}

\altaffiltext{1}{leonard@uwm.edu}
\altaffiltext{2}{wkomp@uwm.edu}
\altaffiltext{3}{vanzella@uwm.edu}

\begin{abstract}

As shown by Parker and Raval, quantum field theory in
curved spacetime gives a possible mechanism for explaining
the observed recent acceleration of the universe. 
This mechanism, which differs in its dynamics from quintessence
models,
causes the universe to make a transition to an
accelerating expansion in which the scalar curvature, $R$, 
of spacetime remains constant. 
This transition occurs despite the fact that we set the renormalized
cosmological constant to zero.
We show that this model 
agrees very well with the current observed type-Ia 
supernova (SNe-Ia) data. There are no free parameters in this fit, 
as the relevant observables are determined independently 
by means of the current cosmic microwave background 
radiation (CMBR) data. We also give the predicted
curves for number count tests and for the ratio, $w(z)$, of 
the dark energy pressure to its density, as well as
for $dw(z)/dz$ versus $w(z)$. These curves differ 
significantly from those obtained from a cosmological 
constant, and will be tested by planned future observations.
\end{abstract}

\keywords{cosmic microwave background ---
cosmological parameters --- 
cosmology: observations ---
cosmology: theory ---
gravitation ---
supernovae: general}

\section{Introduction}
\label{sec:intro}

Observational evidence appears increasingly strong
that the expansion of the universe is undergoing 
acceleration that started at a redshift $z$ of order 1
\citep{Riess98,Riess01,Perlmutter98,Perlmutter99}.
Observations of scores of type-Ia supernovae (SNe-Ia)
out to $z$ of about 1.7 support this 
view~\citep{Riess01}, and even 
glimpse the earlier decelerating stage of the expansion. 
It is fair to say that one of the most important
questions in physics is: what causes this acceleration?

One of the more obvious possible answers is that we
are observing the effects of a small positive 
cosmological constant $\Lambda$ 
\citep{Krauss95,Ostriker95,Dodelson96,Colberg00}.
Another, less obvious possibility is that there is
a quintessence field responsible for the acceleration
of the universe 
\citep{Caldwell98,Zlatev99,Dodelson00,Picon01}. 
Quintessence fields are scalar
fields with potential energy functions that produce
an acceleration of the universe when the gravitational
and classical scalar (quintessence) field equations
are solved. More recently, 
\citet{Parker99a,Parker99b,Parker99c,Parker00,Parker01} 
showed that
a quantized free scalar field of very small mass 
in its vacuum state may accelerate the 
universe.
Their model differs from any quintessence model in that
the scalar field is free, thus interacting only with 
the gravitational field.
The nontrivial dynamics of this model arises from well-defined
finite quantum corrections to the action that appear only in curved 
spacetime.
This was the first model to present a 
realization of dark energy
with ratio of pressure
to energy density taking values more negative than $-1$. Other models
having this property have subsequently been proposed 
\citep{Caldwell02,Melchiorri02}.

The physics of the Parker-Raval model is based on quantum field
theory in curved spacetime. 
The renormalized (i.e., observed) cosmological constant $\Lambda$
is set to zero. Several mechanisms have been proposed that tend to 
drive the value of $\Lambda$ to zero 
\citep{Dolgov83,Ford87,Ford02,Tsamis98a,Tsamis98b,Abramo99},
but these mechanisms play no role in this model.
The energy-momentum
tensor of the quantized field in its vacuum state is determined
by calculating an effective action 
\citep{Schwinger51,DeWitt65,Jackiw74}
in a general curved spacetime. The spacetime is unquantized, and
is itself determined self-consistently from the Einstein
gravitational field equations involving the vacuum expectation
value of the energy-momentum tensor, as well as the classical
energy momentum tensor of matter (including cold dark matter)
and radiation. In solving the Einstein equations,
the symmetries of the FRW spacetime are imposed, but the
spacetime is not otherwise taken as fixed, and the initial
value constraints of general relativity are satisfied. The
acceleration is the result of including in the effective
action a non-perturbative term involving the scalar curvature
of the spacetime
\citep{Parker85a,Parker85b,Parker85c,Jack85}. 
The minimal effective action that includes
this non-perturbative effect and gives the correct trace anomaly
of the energy-momentum tensor was used by Parker and Raval. 
In applying this effective
action to the recent expansion of the universe, terms involving
more than two derivatives of the metric were neglected.
In this approximation, a solution was proposed in which the universe
undergoes a rapid transition from a standard FRW universe dominated
by cold dark matter to one containing significant contributions
of vacuum energy and pressure. The proposal is that this
negative vacuum pressure is responsible for the observed
acceleration of the universe. The reaction back of this negative
vacuum pressure on the expansion of the universe is such as
to cause a rapid transition to an expansion of the universe
in which the scalar curvature remains constant.
The vacuum pressure and energy density are determined by a
single parameter related to the mass of the scalar field.
The transition to constant scalar curvature is the result of
a rapid growth in the magnitudes of the vacuum pressure and
energy density that occurs in this theory when the scalar
curvature approaches a particular value, of the order of the
square of the mass of the particle associated with the scalar
field. The Einstein equations cause a reaction back on the
metric such as to prevent further increase in the magnitudes
of the vacuum pressure and energy density. (This effect is
analogous to Lenz's law in electromagnetism.)

The essential cosmological features of this model may be 
described quite simply.
For times earlier than a time $t_j$ 
(corresponding to $z \sim 1$), the universe
undergoes the stages of the standard model, including
early inflation, and radiation domination followed by 
domination by cold dark matter. 
During the latter stage, at time $t_j$
the vacuum energy and (negative) pressure of 
the free scalar quantized field increase rapidly in
magnitude (from a cosmological point of view). 
The effect of this vacuum 
energy and pressure is to cause the scalar curvature 
$R$ of the spacetime to become constant at a value $R_j$. 
The spacetime line element is that of an FRW universe:
\begin{equation}
ds^2 = -dt^2 + a(t)^2 [(1 - k r^2)^{-1} dr^2 + 
r^2 d\theta^2 + r^2 \sin^{2}{\theta} d\phi^2]\;, \label{ds2}
\end{equation}
where $k= {\pm 1}$ or $0$ indicates the spatial curvature.
By joining at $t_j$
the matter dominated scale parameter $a(t)$ and its 
first and second derivatives to the solution for
$a(t)$ in a constant $R$ universe, one uniquely
determines the scale parameter $a(t)$ for times
after $t_j$. This model is known as the vacuum cold dark matter (VCDM)
model of Parker and Raval, or as the {\em vacuum metamorphosis
model} \citep{Parker99c}
to emphasize the existence of a rapid transition in the
vacuum energy density and pressure.

The constant value, $R_j$, of the scalar curvature is a
function of a single new parameter, $\mbar$, related to
the mass of the free scalar field. Therefore, the function
$a(t)$ for $t > t_j$ is fully determined by $\mbar$. 
The values of $\mbar$ and $t_j$ can 
be expressed in terms of observables, namely, the 
present Hubble constant $H_0$, the densities 
$\Omega_{m0}$ and $\Omega_{r0}$ 
of the matter and radiation, respectively, relative to the 
closure density, and the curvature parameter $\Omega_{k0}
\equiv -k/(H_0^2 a_0^2)$. (Here $a_0 \equiv a(t_0)$ is the present
value of the cosmological scale parameter.)
These observables have been determined
with reasonably good precision by various measurements
that are independent of the SNe-Ia 
\citep{Krauss00,Freedman01,Hu01,Huterer01,Turner01,Wang02}.  
Therefore, the value of $\mbar$ is known to within narrow 
bounds, independently of the SNe-Ia observations.

The power spectrum of the CMBR depends largely on
physical processes occurring long before $t_j$. The
behavior of $a(t)$ in the VCDM model does not significantly 
differ from that of the standard model
until after $t_j$. Therefore, the predicted power spectrum
in the VCDM model differs only slightly from that of the
standard model. We calculate the predicted power spectrum
of the CMBR in the VCDM model (as described below), and
find the range of values of the above observables
that give a good fit to the CMBR observations. 

From this range of observables, the corresponding
range of the parameter $\mbar$ follows. Therefore,
the prediction of the VCDM model for the magnitude versus
redshift curve of the SNe-Ia is completely determined, with
no adjustable parameters. We plot the predicted curves 
(obtained from this range of $\mbar$)
for the distance modulus $\Delta(m - M)$ of the SNe-Ia as 
a function of $z$. Comparison with the observed data points,
as summarized by \citet{Riess01}, shows that a significant
subset of 
predicted
curves fit the SNe-Ia data very well, passing within
the narrow error bars of each of the binned data points,
as well as of the single data 
point at $z \approx 1.7$.

We also give the curves predicted by the VCDM model for 
number counts of cosmological objects 
as a function of $z$ and for the ratio, $w(z)$, 
of the vacuum pressure to vacuum energy density, as well as for
$dw/dz$ versus $w$ (parametrized by $z$). 
The predictions of the VCDM model differ significantly from those
of the $\Lambda$CDM model.
Accurate 
measurements of these quantities out to $z$ of about 2 would 
be very telling.

The model we consider here is the simplest of a class of models in 
which a transition occurs around a finite value of $z$ within the 
range of possible observation. This general class has been studied 
from a phenomenological point of view by 
\citet{Bassett02a,Bassett02b}.
They find that the CMBR, large scale structure, and supernova data 
tend to favor a late-time transition over the standard $\Lambda$CDM 
model. Although they considered only dark-energy equations of state, 
$w$, with $w > -1$, their phenomenological analysis can be 
generalized to include the present VCDM model. In addition, the VCDM 
model is readily generalized to include a nonzero vacuum expectation 
value of the low mass scalar field, which could bring $w$ into the 
range greater than $-1$. In the present paper, we are taking the 
simplest of the possible VCDM (or vacuum metamorphosis) models, so 
as to introduce no arbitrary parameters into the fit to the 
supernova data.

\section{How Observables Determine $\mbar$}
\label{sec:mbar}

In this section, we explain how the value of $\mbar$ 
is obtained from $H_0$, $\Omega_{k0}$, $\Omega_{m0}$, and 
$\Omega_{r0}$ in the VCDM model. (Here, 
$\Omega_{m0} \equiv \Omega_{{\rm cdm}0} + \Omega_{b0}$, 
where $\Omega_{{\rm cdm}0}$ and $\Omega_{b0}$ are the present 
densities of cold dark matter and baryons, respectively, relative to 
the closure density.) 
The relation between $\mbar$ and these present observables
follows from the Einstein equations, the previously described 
constancy of the scalar curvature $R=R_j$ for $t > t_j$, and 
continuity of $a(t)$ and its first and second derivatives 
at time $t_j$. 
For our present purposes, we may define the parameter $\mbar$
in terms of $R_j$, namely, by the relation $\mbar^2 = R_j$.
(At a microscopic level, $\mbar$ is proportional to the mass
of the free quantized scalar field.)

The trace of the Einstein equations at time $t_j$ is
\begin{equation}
\mbar^2 = R_j = 8\pi G \rho_{mj}\; , \label{Rtj}
\end{equation}
where $\rho_{mj}$ is the energy density of the non-relativistic
matter present at time $t_j$. The density and pressure of the
dark energy, $\rho_v$ and $p_v$, respectively, will be taken
to be zero at $t_j$. We make this assumption in order to avoid
introducing a second parameter in addition to $\mbar$.
It should be noted for future reference that dropping this 
assumption will introduce another parameter that would affect
mainly the behavior of the predicted SNe-Ia curve near the
transition time $t_j$ from matter-dominated to 
constant-scalar-curvature universe. In the present paper, we 
do not relax this assumption of zero $\rho_{vj}$ and $p_{vj}$ 
because we find good agreement of the one parameter VCDM 
model with the current observational data.

For all $t > t_j$, the scalar curvature is taken to remain
constant at the value $R_j = \mbar^2$. Thus,
\begin{equation}
6[(\dot a/a)^2 + (\ddot a/a) + k/a^2] = \mbar^2\; ,
\label{Rt}
\end{equation}
where dots represent time derivatives.
Defining the variable $x \equiv a^2$, this becomes
\begin{equation}
\frac{1}{2}\ddot x +  k = \frac{1}{6}\mbar^2 x \; .
\label{oscillator}
\end{equation}
The first integral of this equation is
\begin{equation}
\frac{1}{4}\dot{x}^2 - \frac{1}{12}\mbar^2 x^2 + k x = E \;,
\label{energy}
\end{equation}
where $E$ is a constant.

One of the Einstein equations at time $t_j$ is
\begin{equation}
H_j^2 + k/a_j^2 = (8\pi G/3) \rho_j\;, 
\label{Einsteina}
\end{equation}
where $H_j\equiv {\dot a}(t_j)/a(t_j)$, $a_j \equiv a(t_j)$, 
and $\rho_j\equiv \rho(t_j)$ is the total energy density at 
time $t = t_j$.
The last equation can be rewritten as
\begin{equation}
\frac{1}{4}\dot{x}_j^2 + k x_j =
  (8\pi G/3) \rho_j x_j^2\;,
\label{Einsteinx}
\end{equation}
where subscript $j$ refers to quantities at time $t_j$.
Comparing this with equation (\ref{energy}), one finds that
\begin{equation}
E = \left[(8\pi G/3) \rho_j - 
                        \frac{1}{12} \mbar^2\right]x_j^2\;.
\label{E1}
\end{equation}
Using $\rho_j = \rho_{mj} + \rho_{rj}$, where $\rho_{rj}$ is
the radiation energy density at time $t_j$, and
equation (\ref{Rtj}) to eliminate $\rho_j$ and $\rho_{mj}$
from the last expression for $E$, we find that
\begin{equation}
E = (8\pi G/3) \rho_{rj} x_j^2 + \frac{1}{4} \mbar^2 x_j^2\;.
\label{E2}
\end{equation}
Thus, equation (\ref{energy}) gives the following conserved quantity:
\begin{equation}
\frac{1}{4}\dot{x}^2 - \frac{1}{12}\mbar^2 x^2 + k x = 
(8\pi G/3) \rho_{rj} x_j^2 + \frac{1}{4} \mbar^2 x_j^2\;.
\label{E3}
\end{equation}
This is readily written in terms of $a(t)$ and its derivatives.
We have $\dot x/ x = 2 \dot a/ a \equiv 2 H(t)$, and
$ x_j^2/x(t)^2 = [a_j/a(t)]^4$. Hence, the radiation energy
density satisfies
\begin{equation}
 \rho_{rj} x_j^2/x(t)^2 = \rho_r(t)\;,
\label{radiation}
\end{equation}
and equation (\ref{E3}) is
\begin{equation}
H(t)^2 + k/a(t)^2 =  
   (8\pi G/3) \rho_r(t) + \frac{1}{4} \mbar^2 [a_j/a(t)]^4
      + \frac{1}{12}\mbar^2\;.
\label{E4}
\end{equation}
Solving for $[a_j/a(t)]^4$, we obtain (for $t \geq t_j$),
\begin{equation}
[a_j/a(t)]^4 = 
\frac{4}{\mbar^2}\left[ H(t)^2
   + k/a(t)^2 - (8\pi G/3) \rho_r(t)\right] -\frac{1}{3} \;.
\label{a4ratio}
\end{equation}

Returning to equation (\ref{Rtj}), we can now express $\mbar$
in terms of the present values 
of $\rho_m$, $\rho_r$, $H$, and $k/a$.
We have
\begin{equation}
\mbar^2 = R_j = 8\pi G \rho_{m0} (a_0/a_j)^3\;.
\label{Rtj2}
\end{equation}
With equation (\ref{a4ratio}), it follows that
\begin{equation}
\mbar^2 = 8\pi G \rho_{m0}
  \left\{\frac{4}{\mbar^2}\left[ H_0^2
   + k/a_0^2 - (8\pi G/3) \rho_{r0} \right] 
   -\frac{1}{3}\right\}^{-3/4} .
\label{m2}
\end{equation}
Using the expression for the present critical density, 
$\rho_{c0} = 3 H_0^2 / (8\pi G)$, equation (\ref{m2})
takes the dimensionless form,
\begin{equation}
{m_{H0}}^2 = 3 \Omega_{m0}
  \left\{ (4/m_{H0}^2)\left[
   1 - \Omega_{k0} - \Omega_{r0}\right] 
   -\frac{1}{3}\right\}^{-3/4},
\label{mHdimensionless}
\end{equation}
where $m_{H0} \equiv \mbar / H_0$. 
This is readily solved numerically for $m_{H0}^2$.
Alternatively, it can be put into the form of a 
fourth-order equation for $m_{H0}^2$ and solved analytically. 
Using the values $\Omega_{m0}=0.34^{+0.46}_{-0.14}$ and 
$H_0=65^{-16}_{+10}$~${\rm km}~{\rm s}^{-1}~{\rm Mpc}^{-1}$
obtained in section~\ref{sec:CMBR} from the CMBR power spectrum 
(see fig.~\ref{fig:95cl}),
as well as
$\Omega_{r0}=8.33\times 10^{-5}$ and $\Omega_{k0}=0$, we
find 
$
m_{H0}=3.26^{-0.62}_{+0.10}
$
and
$
\mbar=4.5^{-1.7}_{+0.9}\times 10^{-33}\;\;{\rm eV}
$.
(The uncertainties
refer to the $95\%$ confidence level.) Note that the fitting of the 
CMBR
power spectrum alone gives us a wide range for $\Omega_{m0}$ and 
$H_0$. 
In order 
to further restrict our results (thus being able to make stronger 
predictions),
we will adopt the HST-Key-Project result
$H_0=72 
\pm 8$~${\rm km}~{\rm s}^{-1}~{\rm Mpc}^{-1}$~\citep{Freedman01}
as a constraint. 
This narrows the range of our cosmological parameters down to
$\Omega_{m0}=0.34^{+0.08}_{-0.14}$ and 
$H_0=65^{-1}_{+10}$~${\rm km}~{\rm s}^{-1}~{\rm Mpc}^{-1}$, as can 
be seen
from figure~\ref{fig:95cl}. Aiming at combining these two methods of 
determining the uncertainties of
our results, hereafter we adopt the notation exemplified by
$\Omega_{m0}=0.34^{+(0.46;0.08)}_{-0.14}$ and 
$H_0=65^{-(16;1)}_{+10}$~${\rm km}~{\rm s}^{-1}~{\rm Mpc}^{-1}$, where
the uncertainties appearing in parenthesis refer to the $95\%$ 
confidence level, without and with
the HST constraint, respectively. (The sign appearing in 
front of the 
parenthesis is common to both uncertainties.) Thus, returning to the
parameters $m_{H0}$ and $\mbar$, we have:
\begin{equation}
m_{H0}=3.26^{-(0.62;0.08)}_{+0.10}\;
\label{mH0value}
\end{equation}
and
\begin{equation} 
\mbar=4.52^{-(1.76;0.18)}_{+0.84}\times 10^{-33}\;\;{\rm eV}\;. 
\label{mbarvalue}
\end{equation}

Finally, the solution to equation (\ref{oscillator}) 
for $x(t) = a(t)^2$ is 
\begin{eqnarray}
a(t)^2/a_0^2 &=& 
     \cosh\left(\frac{\mbar}{\sqrt{3}} (t-t_0)\right)\nonumber\\
     & & +\frac{2 \sqrt{3}}{m_{H0}} 
        \sinh\left(\frac{\mbar}{\sqrt{3}} (t-t_0)\right)\nonumber\\
     & & -\frac{6 \Omega_{k0}}{m_{H0}^2} 
        \cosh\left(\frac{\mbar}{\sqrt{3}} (t-t_0)\right)
       +\frac{6 \Omega_{k0}}{m_{H0}^2}\;.
\label{solx}
\end{eqnarray}
It then follows from $(1/2)\dot{x}(t) / x(t) = H(t)$, that
\begin{eqnarray}
H(t)/H_0 &=& a(t)^{-2} a_0^2 \left[
  \cosh\left(\frac{\mbar}{\sqrt{3}} (t-t_0)\right) \right.
\nonumber\\
    & & \left. +\sqrt{3}\left(\frac{m_{H0}}{6} - 
          \frac{\Omega_{k0}}{m_{H0}}\right) 
          \sinh\left(\frac{\mbar}{\sqrt{3}} (t-t_0)\right)
         \right]\;.
\label{solH}
\end{eqnarray}

From 
equations~(\ref{E4}), (\ref{Rtj2}), and (\ref{mHdimensionless})
one obtains the following expression in terms of the
redshift, $z \equiv a_0/a - 1$:
\begin{equation}
H(z)^2/H_0^2  =  
    (1 - \Omega_{k0} - m_{H0}^2/12)(1+z)^4 +
      \Omega_{k0}(1+z)^2 + m_{H0}^2/12\;.
\label{Hofz}
\end{equation}
From equations~(\ref{Rtj2}) and (\ref{mH0value}) 
one finds the redshift $z_j$
at time $t_j$,
\begin{equation}
z_j  = [m_{H0}^2/(3 \Omega_{m0})]^{1/3}-1=
1.19^{-(0.76;0.19)}_{+0.47} \;.
\label{zj}
\end{equation}
Moreover, from equation~(\ref{Hofz}) we can obtain the redshift, 
$z_a$, at which the expansion of the Universe starts to accelerate.
In fact, by looking at the deceleration parameter,
\begin{eqnarray}
q \equiv -\frac{a \ddot{a}}{\dot{a}^2} =
(1+z) \left[\ln\left(H/H_0\right)\right]'-1\;,
\label{decelerating}
\end{eqnarray}
where the prime sign stands for derivative with respect to $z$, 
we have that
$z_a$ satisfies 
\begin{equation}
(1+z_a)H'(z_a)=H(z_a)\;.
\label{Hza}
\end{equation}
Thus, from equation~(\ref{Hofz}) and the cosmological parameters
mentioned above, we obtain for the spatially flat VCDM model
\begin{equation}
z_a=\left[
\frac{m_{H0}^2/12}{1-\Omega_{k0}-m_{H0}^2/12}
\right]^{1/4}-1=0.67^{-(0.59;0.15)}_{+0.35}\;.
\label{za}
\end{equation}
This value is similar to the one obtained using
the spatially flat 
$\Lambda$CDM model with $\Omega_{\Lambda 0}=0.67$:
\begin{equation}
z_{a \Lambda{\rm CDM}}\approx
\left(
\frac{2\Omega_{\Lambda 0}}{1-\Omega_{\Lambda 0}}
\right)^{1/3}-1\approx 0.60\;.
\label{zalcdm}
\end{equation}

\section{The Dark Energy}
\label{sec:de}

In the VCDM model, the dark energy is the energy
of the vacuum, denoted by $\rho_v$. 
This vacuum energy is not in the form of real 
particles, but may be thought of as energy associated 
with fluctuations (or virtual particles) of the 
quantized scalar field. Vacuum energy, $\rho_v$, 
and pressure, $p_v$, must be included as a source of 
gravitation in the Einstein equations. 
Thus, for $t>t_j$, one has
\begin{equation}
H(t)^2 + k/a(t)^2 =  
   (8\pi G/3) \left[\rho_r(t) + \rho_m(t) + \rho_v(t)\right]\;.
\label{Einsteint}
\end{equation}

The vacuum energy and pressure remain essentially zero 
until the time $t_j$ when the value of the scalar 
curvature $R$ has fallen to a value slightly greater 
than $\mbar^2$. Then in a short time (on a cosmological scale),
the vacuum energy and pressure grow, and through
their reaction back cause the scalar curvature to remain
essentially constant at a value just above 
$R_j = \mbar^2$~\citep{Parker99a,Parker99b}.
Intuitively, this reaction
back may be thought of as similar to what happens in
electromagnetism when a bar magnet is pushed into a coil
of wire. The current induced in the coil produces a
magnetic field that opposes the motion of the bar magnet
into the coil (Lenz's Law). Similarly, in the present
case, the matter dominated expansion of the universe
causes the scalar curvature to decrease. But as it
approaches the critical value, $\mbar^2$, the
quantum contributions to the energy-momentum tensor
of the scalar field grow large in such a way as to oppose 
the decrease in $R$ that is responsible for the growth in
quantum contributions. The universe continues to expand,
but in such a way as to keep $R$ from decreasing further.

Defining $t_j$ as the time at which $\rho_v$ and $p_v$
begin to grow significantly, we have to good approximation,
equation (\ref{Rtj}). Evolving $\rho_{m_j}$ forward in time, 
then gives
\begin{equation}
8 \pi G \rho_m(t) = \mbar^2 \left[ a_j/a(t) \right]^3.
\label{rhom}
\end{equation}
One then finds from equation (\ref{Einsteint}) and equation (\ref{E4})
that the vacuum energy density evolves for $t>t_j$ as
\begin{equation}
\rho_v(t) = \frac{\mbar^2}{32 \pi G} \left\{
       1 -
       4 \left[a_j/a(t)\right]^3 +
       3 \left[a_j/a(t)\right]^4
\right\}\;.
\label{rhov}
\end{equation}
The conservation laws for the total energy density and pressure
and for the energy densities and pressures of the 
radiation alone and of the matter alone, then 
imply that $\rho_v$ and the vacuum pressure, $p_v$, 
also satisfy the conservation law. It follows that 
\begin{eqnarray}
p_v(t) &=& -\frac{d}{dt}(\rho_v a^3)/ \frac{d}{dt}(a^3)\nonumber\\
    &=& \frac{\mbar^2}{32 \pi G}\left\{ -1 + 
              \left[ a_j/a(t)\right]^4 \right\} \;.
\label{pv}
\end{eqnarray}
At late times, as $a_j/a(t)$ approaches zero, one sees that
the vacuum energy density and pressure approach those of a
cosmological constant, $\Lambda = \mbar^2 /4$. But at 
finite times, their time evolution differs from that of 
a cosmological constant.

One immediately sees from equations~(\ref{rhom}) and (\ref{rhov})
that for $t>t_j$,
\begin{equation}
\rho_v(t)+\rho_m(t) = \frac{\mbar^2}{32 \pi G} \left\{
       1 +
       3 \left[a_j/a(t)\right]^4
\right\}\;.
\label{rhovrhom}
\end{equation}
Using equation~(\ref{pv}), it now follows that
\begin{equation}
p_v(t) = (1/3)\left[\rho_v(t)+\rho_m(t)\right]
          -  \mbar^2/(24\pi G)\;.
\label{pvrhovrhom}
\end{equation}
Since $p_m = 0$, and $p_r = (1/3) \rho_r$, the total
pressure, $p$, and energy density, $\rho$, satisfy the
equation of state~\citep{Parker00}
\begin{equation}
p(t) = (1/3)\rho(t) -  \mbar^2/(24\pi G)\;.
\label{prhoeqofstate}
\end{equation}
From this equation of state and the conservation law,
which can be written in the form,
$d(\rho a^3)/da + p d(a^3)/da = 0$, we find that
for $t>t_j$,
\begin{equation}
\rho(a) = 
\left(\rho_j -\frac{\mbar^2}{32\pi G}\right)
(a_j/a)^4 + \frac{\mbar^2}{32 \pi G}\;.
\label{rhoa1}
\end{equation}
As the vacuum energy density is
taken as zero at $t=t_j$, we have $\rho_j = \rho_{mj}
+ \rho_{rj} = \rho_{m0} a_0^3/a_j^3 +
\rho_{r0} a_0^4/a_j^4$, and then, from equation~(\ref{Rtj2}),
\begin{equation}
\rho_j = \mbar^2/(8\pi G)
       +\rho_{r0}  \left[\mbar^2/(8\pi G \rho_{m0})\right]^{4/3}.
\label{rhoa2}
\end{equation}
Then, with equation~(\ref{rhom}) and (for $t>t_j$) 
$(a_j/a)^4 = (a_0/a)^4  ( 8\pi G \rho_{m0}/\mbar^2)^{4/3}$,
equation~(\ref{rhoa1}) finally becomes
\begin{equation}
\rho(a) = \left[
\rho_{m0}\, (3\Omega_{m0}/m_{H0}^2)^{1/3} +\rho_{r0}\right]\, 
(a_0/a)^4  
   +  \mbar^2/(32\pi G)\;.
\label{rhoa4}
\end{equation}
This expression will be used in the next section
to calculate, among other things,
the age of the Universe as predicted by the VCDM model.

\section{Age of the universe}
\label{sec:au}

The values of $t_j$ and $t_0$ are found by integration:
\begin{equation}
t = \int_0^t\, dt = \int_0^{a(t)} \, da \, a^{-1} H(a)^{-1}.
\label{tintegral1}
\end{equation}
From equation (\ref{Einsteint}), 
\begin{equation}
t =  \int_0^{a(t)}\, da \, a^{-1} \, 
\left\{-k/a^2 + (8\pi G/3) 
\left[\rho_r(a)+\rho_m(a)+\rho_v(a)\right]
\right\}^{-1/2},
\label{tintegral2}
\end{equation}
This integral is conveniently split in two at time $t_j$,
and expressed in terms of variable of 
integration $y \equiv a/a_0$: 
\begin{equation}
t_j =  (H_0)^{-1} \int_0^{a_j/a_0}\, dy \, 
 \left(\Omega_{k0} + \Omega_{m0} y^{-1}+\Omega_{r0} y^{-2}
\right)^{-1/2},
\label{tjintegral}
\end{equation}
and from equation (\ref{rhoa4}),
\begin{equation}
t_0 - t_j =  (H_0)^{-1} \int_{a_j/a_0}^{1}\, dy \, 
\left\{ \Omega_{k0} + \left[
\Omega_{m0}\, (3\Omega_{m0}/m_{H0}^2)^{1/3} +\Omega_{r0}\right]\, 
y^{-2}  
   +  (m_{H0}^2/12)\, y^2 \right\}^{-1/2}.
\label{tjt0integral2}
\end{equation}
Another interesting parameter to obtain is $t_a$, the
time when the expansion of the Universe starts to accelerate:
\begin{equation}
t_0-t_a=(H_0)^{-1} \int_{a_a/a_0}^{1}\, dy \, 
\left\{ \Omega_{k0} + \left[
\Omega_{m0}\, (3\Omega_{m0}/m_{H0}^2)^{1/3} +\Omega_{r0}\right]\, 
y^{-2}  
   +  (m_{H0}^2/12)\, y^2 \right\}^{-1/2},
\label{t0taintegral}
\end{equation}
where $a_a\equiv a(t_a)$.

Using the cosmological parameters obtained in 
section~\ref{sec:CMBR} by fitting the VCDM model
to the CMBR power spectrum data, 
we have (see eqs.~[\ref{zj}] and [\ref{za}])
$
a_j/a_0=(1+z_j)^{-1}=0.46^{+(0.24;0.04)}_{-0.08}
$$\;$
and$\;$
$
a_a/a_0=(1+z_a)^{-1}= 0.60^{+(0.32;0.06)}_{-0.10}
$,$\;$
and$\;$ consequently,$\;$
$H_0 t_j=0.354^{+(0.084;0.009)}_{-0.011}$, 
$H_0t_a = 0.535^{+(0.126;0.013)}_{-0.017}$, and
$H_0 t_0= 0.99^{-(0.26;0.07)}_{+0.16}$.
Thus, using the correspondent values of $H_0$,
we finally obtain
\begin{equation}
t_j=5.33^{+(3.41;0.22)}_{-0.84}\;{\rm Gyr}\;,
\label{tjvalue}
\end{equation}
\begin{equation}
t_a=8.0^{+(5.2;0.4)}_{-1.2} \; {\rm Gyr}\;,
\label{tavalue}
\end{equation}
and
\begin{equation}
t_0=14.9\mp 0.8\;{\rm Gyr}\;.
\label{t0value}
\end{equation}
Note that the uncertainty in $t_0$ is independent of whether or not
we adopt 
the HST-Key-Project result as a constraint. This is 
because the dependence of $t_0$
on $H_0$ and $\Omega_{m0}$ is such that the
$95\%$ confidence region shown in figure~\ref{fig:95cl} happens to be
stretched along lines of constant values of $t_0$. The same
is true for the $\Lambda$CDM model, as discussed by 
\citet{Knox01}.

We can compare the values presented in equations~(\ref{tavalue}) and
(\ref{t0value}) with the respective ones given by the spatially
flat $\Lambda$CDM model.
 Using $\Omega_{\Lambda 0}=0.67$, the $\Lambda$CDM model
gives
$H_0 t_a\approx 0.536$ and 
$H_0 t_0\approx 0.938$.
With the best-fit value of the Hubble constant obtained for our model,
$H_0=65$~${\rm km}~{\rm s}^{-1}~{\rm Mpc}^{-1}$ 
(see section~\ref{sec:CMBR}),
this gives
$t_a\approx 8.1\;{\rm Gyr}$ and
$t_0\approx 14.1\;{\rm Gyr}$. Therefore, 
we see that the age attributed to the Universe by the VCDM model
is larger than the age predicted by the $\Lambda$CDM model, for
essentially the same values of $\Omega_{m0}$ and $H_0$.

\section{Fit to the CMBR power spectrum}
\label{sec:CMBR}

In view of recent CMBR 
observations~\citep{Pryke02,Masi02,Netterfield02,Abroe01}, there is a
need to reexamine the results obtained by
\citet{Parker01}.
In this section, we will obtain the cosmological parameters
$\omega_{{\rm cdm}0}\equiv \Omega_{{\rm cdm}0} h^2$, 
$\omega_{b0} \equiv \Omega_{b0} h^2$, 
and 
$H_0\equiv 100 h$~${\rm km}~{\rm s}^{-1}~{\rm Mpc}^{-1}$
which give the best 
fit of the spatially flat VCDM model to the recent measurements
of the CMBR power spectrum by Boomerang, {\scriptsize MAXIMA}, and 
{\scriptsize DASI}
($h$ is a dimensionless quantity defined by the latter expression). 
As we have 
seen, 
these parameters are the essential ingredients necessary to fix the 
parameter 
$\mbar$. [The other necessary parameter, $\Omega_{r0}=8.33\times 
10^{-5}$, 
is independently obtained from the CMBR mean temperature and the 
number of relic neutrino species~\citep{Peebles93}.]

In order to obtain the CMBR power spectrum fluctuations predicted
by the VCDM model, with given values of $\omega_{{\rm cdm}0}$, 
$\omega_{b0}$, and $H_0$, we use a slightly modified version of the 
{\scriptsize 
CMBFAST}\footnote{http://physics.nyu.edu/matiasz/CMBFAST/cmbfast.html}
computer code~\citep{Seljak96,Zaldarriaga00}.
The modifications made in the code, described by
\citet{Parker01},
consist of adding the vacuum contributions, $\rho_v$ and
$p_v$,
to the total energy density and pressure, respectively,
for time $t>t_j$, i.e., after the transition to dark energy dominated 
epoch.

We set up the {\scriptsize CMBFAST} code to
generate a numerical grid in the 3 dimensional cosmological parameter 
space 
($\omega_{{\rm cdm}0}$, $\omega_{{\rm b}0}$, $H_{0}$).  We
introduce prior information on the value of the present day Hubble 
constant, $H_{0}$, to be in the range from 45 to 80 with units of 
km Mpc$^{-1}$ s$^{-1}$.  
Also, we set the value of the cosmological constant, 
$\Lambda$, to be
zero.  To perform our numerical analysis
consistent with these priors, we generated a class of VCDM
models with the following cosmological parameters and
resolutions (in the form of ``initial value'':``final value'':``step 
size''):
$\omega_{b0}=(0.005:0.030:0.001)$, 
$\omega_{{\rm cdm}0}=(0.05:0.31:0.01)$, 
and 
$H_{0}=(45.0:80.0:1.0)$~${\rm km}~{\rm s}^{-1}~{\rm Mpc}^{-1}$. 
We chose to vary these
three parameters based on the fact that they determine directly
$\bar{m}$, which is the one free parameter of the VCDM model.  All
models generated use the Radical Compression Data Analysis 
Package\footnote{http://bubba.ucdavis.edu/$\sim$knox/radpack.html}
(RadPack) \citep{Bond00} to compute a $\chi^{2}$ test
statistic that compares the predicted
CMBR spectrum to the experimental
measurements of  
{\scriptsize DASI} \citep{Pryke02},
Boomerang \citep{Masi02,Netterfield02}, and 
{\scriptsize MAXIMA} 
\citep{Abroe01} at
particular multipoles $l$.  We look for minima of $\chi^{2}$
in the class of cosmologies specified above.  The particular
VCDM model described by the
parameters which
give the
minimum of $\chi^{2}$ in our parameter space is
called the best fit model.
\begin{deluxetable}{ccccc}
\tablewidth{0pt}
\tablecaption{Multipole numbers $l$ and associated power intensities
$I_l$ of the
first few peaks and troughs of the CMBR spectrum predicted by
the spatially flat 
VCDM model (see fig.~\ref{fig:CMBR}). The
uncertainties $\Delta l$ and
$\Delta I_l$ are calculated from the $\chi^{2}$ test
statistic. 
\label{tab:peaks}}
\tablehead{
\colhead{}                          &
\colhead{$l$}                 & \colhead{$\Delta l$}                &
\colhead{$I_l\;(\mu {\rm K}^2)$}  & 
\colhead{$\Delta I_l\;(\mu {\rm K}^2)$}}
\startdata
Peaks   & $230$ & $\pm 1$  & $5133$ & $+22$/$-337$ \\
        & $558$ & $+7$/$-5$  & $2436$ & $+162$/$-123$ \\
        & $848$ & $+4$/$-9$ & $2441$ & $+291$/$-229$   \\
\\
Troughs & $427$ & $+2$/$-7$  & $1591$ & $+132$/$-81$  \\
        & $700$ & $+7$/$-20$ & $1685$ & $+173$/$-158$  \\
\enddata
\end{deluxetable}
\begin{figure}
\plotone{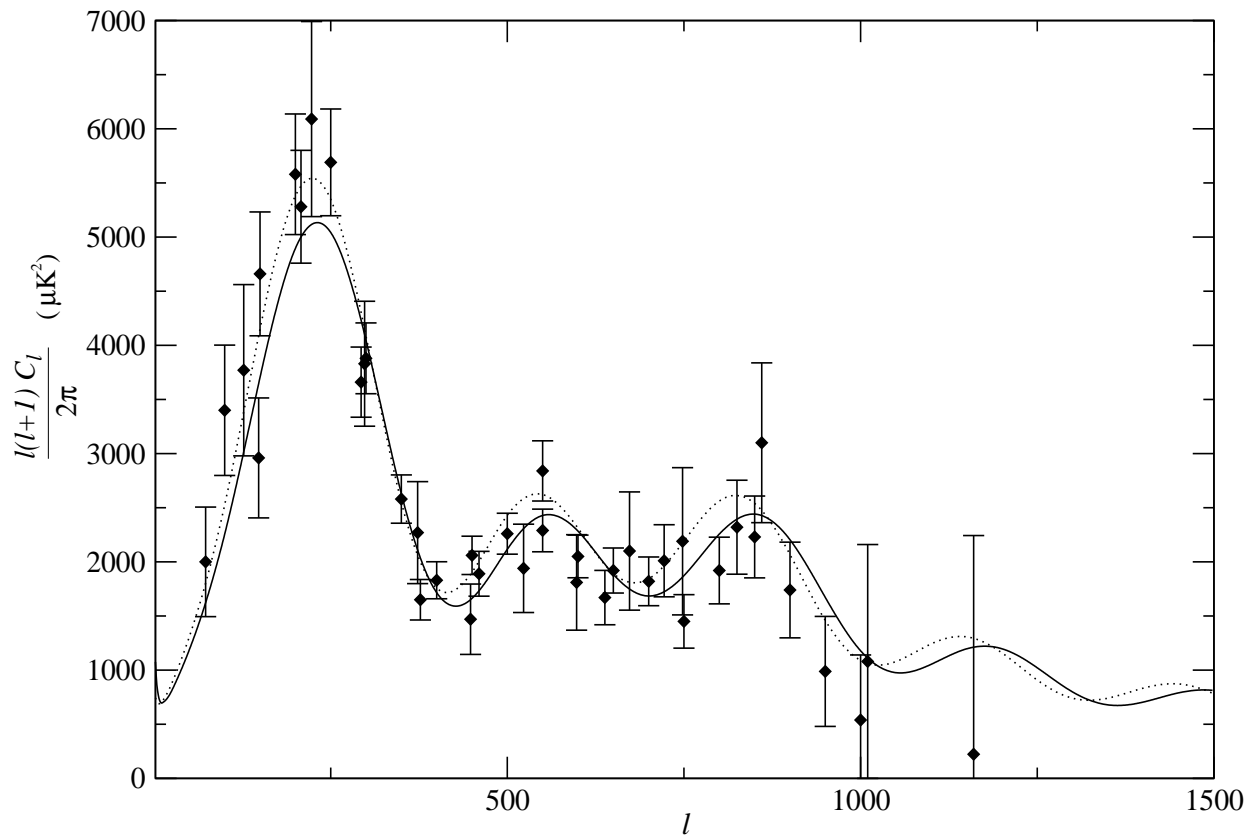}
\caption{Plot of the best fit CMBR
power spectrum for the spatially flat
VCDM model (solid curve) and of the CMBR power spectrum for the
spatially flat
$\Lambda$CDM model with $\Omega_{\Lambda 0}=0.67$ (dashed curve).
The diamond points and error bars correspond to {\scriptsize DASI}, 
Boomerang,
and {\scriptsize MAXIMA} experimental data.}
\label{fig:CMBR}
\end{figure}

For the best fit VCDM model, we found $\omega_{b0}=0.022$,
$\omega_{{\rm cdm}0}=0.12$, and 
$H_{0}=65.0$ ${\rm km}$ ${\rm s}^{-1}$
${\rm Mpc}^{-1}$. This best fit has
$\chi^2=\chi_{min}^{2}=48.84$, corresponding to a significance level
$\alpha(\chi_{min}) 
\equiv \int_{\chi_{min}^2}^{\infty} f(\chi^2,n)\;d(\chi^2)
=0.187$, where
$f(\chi^2,n)=\left[2\Gamma(n/2)\right]^{-1}
(\chi^2/2)^{n/2-1}e^{-\chi^2/2}$ is the $\chi^2$ probability
density function (pdf) for $n$ degrees of freedom
(the recent CMBR data we consider 
are constituted of $41$ experimental points).
The three
best fit cosmological parameters of the VCDM model are
similar to the values obtained from the $\Lambda$CDM
model fit to the CMBR power 
spectrum~\citep{Pryke02,Abroe01,Masi02,Netterfield02}.

To find the 95$\%$ confidence region of our parameter space, 
we compute 
the quantity
$\chi_{95\%}^{2}$ where
$\alpha(\chi_{95\%})=0.05$, 
and then look for the subset of VCDM models 
with $\chi^{2} \le \chi^{2}_{95\%}$ which lie in  the parameter 
space given above.  
The results for the VCDM model are 
$H_{0}=65_{+10}^{-16}$~${\rm km}~{\rm s}^{-1}~{\rm Mpc}^{-1}$, 
$\omega_{{\rm cdm}0}=0.12^{+0.06}_{-0.03}$, and 
$\omega_{b0}=0.022^{+0.003}_{-0.004}$.
The uncertainties given above are  obtained from the extrema
of the 95$\%$ confidence region.  
The range of $\omega_{b0}$ is 
consistent with Big Bang Nucleosynthesis~\citep{Bean02},
and the best fit value for $H_0$ lies within the range given by the
HST Key Project~\citep{Freedman01}.

\begin{figure}
\plotone{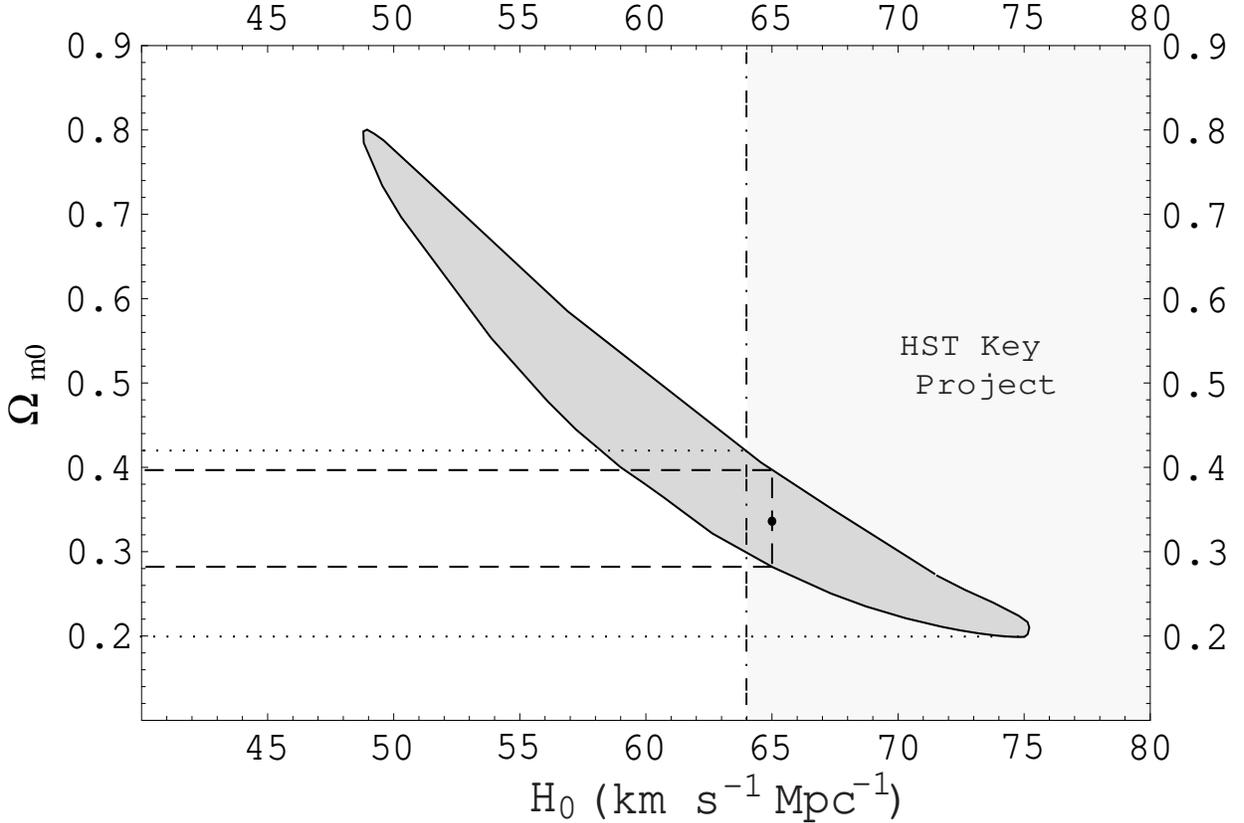}
\caption{Plot of the $95\%$ confidence level (dark-gray region) 
in the parameters $\Omega_{m0}$ and $H_0$ obtained by fitting the 
CMBR power spectrum using the VCDM model. The extrema of this
region, together with the best-fit values, define 
the ranges $\Omega_{m0}=0.34^{+0.46}_{-0.14}$ and
$H_0=65^{-16}_{+10}$~${\rm km}~{\rm s}^{-1}~{\rm Mpc}^{-1}$. 
Adopting the HST-Key-Project result 
($H_0=72\pm 8$~${\rm km}~{\rm s}^{-1}~{\rm Mpc}^{-1}$,
represented by the light-gray region), the range in
$\Omega_{m0}$ is narrowed down to $\Omega_{m0}=0.34^{+0.08}_{-0.14}$
(dotted lines). 
Fixing $H_0=65$~${\rm km}~{\rm s}^{-1}~{\rm Mpc}^{-1}$, we have
$\Omega_{m0}=0.34\pm 0.06$ (dashed lines).}
\label{fig:95cl}
\end{figure}

From the $95\%$ confidence region in the parameter space 
($\omega_{{\rm cdm}0}$, $\omega_{b0}$, $H_0$) we can obtain
the $95\%$ confidence region in the space ($\Omega_{m0}$, $H_0$), 
recalling that $\Omega_{m0}=(\omega_{{\rm cdm}0}+\omega_{b0})/h^2$
and $h\equiv H_0/\left(100\;{\rm km}\;{\rm s}^{-1}\;
{\rm Mpc}^{-1}\right)$. This confidence region is presented in
figure~\ref{fig:95cl} (dark-gray region). 
The reason to concentrate in the parameter 
space
($\Omega_{m0}$, $H_0$) is that all quantities we are interested in
(e.g., the age 
of the Universe, the luminosity distances of SNe-Ia, number 
count tests, the dark energy equation of state) depend directly on
$\Omega_{m0}$, not on 
$\omega_{{\rm cdm}0}$ and $\omega_{b0}$ separately (except, obviously,
the quantities directly related to the CMBR power spectrum). From 
figure~\ref{fig:95cl} we see that $\Omega_{m0}=0.34^{+0.46}_{-0.14}$. 
This very wide range is just a consequence of the fact that the 
Hubble constant $H_0$ is not tightly constrained by the CMBR power
spectrum. However, adopting the HST-Key-Project result 
$H_0=72 \pm 8$~${\rm km}~{\rm s}^{-1}~{\rm Mpc}^{-1}$ as a constraint
(light-gray region in fig.~\ref{fig:95cl}), the acceptable
range of $\Omega_{m0}$ can be narrowed down to 
$\Omega_{m0}=0.34^{+0.08}_{-0.14}$, which then 
allows us to make stronger predictions about the cosmological 
quantities mentioned above. Using the even stronger prior
$H_0=65$~${\rm km}~{\rm s}^{-1}~{\rm Mpc}^{-1}$ (the best-fit value)
we obtain the ``best-fit range'' $\Omega_{m0}=0.34 \pm 0.06$,
which will be used in the next section to fit (with no free 
parameters) the luminosity distances of SNe-Ia.

For the CMBR power spectrum predicted by the
best fit flat VCDM model (see fig.~\ref{fig:CMBR}), the multipole 
numbers $l$ and 
power intensities $I_l\equiv l(l+1) C_l/(2\pi)$ of the first three 
peaks and two troughs
are given in table~\ref{tab:peaks}.  
The uncertainties $\Delta l$ and $\Delta I_l$ 
correspond to the 95\% confidence region of $\omega_{{\rm cdm}0}$, 
$\omega_{b0}$, and
$H_{0}$. Comparing the values given in table~\ref{tab:peaks} with
the correspondent results presented by \citet{Durrer01}, 
we see that our values for the first three peaks and two troughs
are well within  the ($1 \sigma$) ranges for these quantities that
can be obtained, in a model-independent way, from the combined
Boomerang, {\scriptsize MAXIMA}, and {\scriptsize DASI} data. For 
instance, the last two columns of 
table~2 of \citet{Durrer01} give the multipole number and the
power intensity of the first peak to be
$l_{p_1}=213^{+35}_{-59}$ and 
$I_{p_1}=5041^{+1017}_{-1196}$~$\mu$K$^2$,
respectively.

The results of this section show that the VCDM model gives a 
reasonable fit to the CMBR power spectrum with values of 
$\omega_{{\rm cdm}0}$, 
$\omega_{b0}$, and
$H_{0}$ that are consistent with current observations.
The future of CMBR
observations looks very promising with a mixture of ground based
interferometers ({\scriptsize DASI}), airborne interferometers 
({\scriptsize MAXIMA} and
Boomerang), and satellite experiments 
(Microwave Anisotropy Probe
 and the Planck satellite) that will further probe
the CMBR anisotropies at higher and lower multipoles.

\section{No-parameter fit to the SNe-Ia data}
\label{sec:SNe}

In the present section we compare the luminosity distance as a 
function of redshift
predicted by the VCDM model to the measured values
of the luminosity distances of SNe-Ia as summarized by 
\citet{Riess01}.
Because all the relevant parameters of the model are determined 
by fitting the CMBR power spectrum 
(see sec.~\ref{sec:CMBR}), this comparison
is a no-parameter fit of the VCDM model to the SNe-Ia data.

We start by computing the luminosity distance as a function of
redshift, $d_L(z)$.
As a consequence of equation~(\ref{ds2}),
the comoving coordinate distance $r(z)$ of objects
observed with redshift $z$ satisfies
\begin{equation}
\frac{dr}{\sqrt{1-kr^{2}}}=
\frac{dt}{a(t)}=
\frac{dz}{a_{0}H(z)}\; ,
\label{drdz}
\end{equation}
which leads to
\begin{eqnarray}
r(z)=
\cases{
\int_0^z\left[a_0H(z')\right]^{-1}dz'&, $k=0$ \cr
\left(a_0H_0 \Omega_{k0}^{1/2}\right)^{-1}
\sinh\left(
H_0 \Omega_{k0}^{1/2}\int_0^z H(z')^{-1}dz'
\right)&, $k\neq 0$},
\label{rz}
\end{eqnarray}
where $H(z)$ for the VCDM model is given by (see eq.~[\ref{Hofz}])
\begin{eqnarray}
\frac{H(z)}{H_{0}} =
\cases{
\left[ \left(1-\Omega_{k0}-
m_{H0}^{2}/12 \right) \left( 1+z \right)^{4} + 
\Omega_{k0} \left(1+z\right)^{2} + m_{H0}^{2}/12 
\right]^{1/2} &,
$z < z_{j}$ \cr
\left[ \Omega_{r0} \left( 1+z \right)^{4}+\Omega_{m0} 
\left( 1+z \right)^{3} + \Omega_{k0} \left( 1+z \right)^{2} 
\right]^{1/2} &,
$z \ge z_{j}$}.
\label{Hz}
\end{eqnarray}

The luminosity distance and the distance modulus are defined
respectively as
\begin{equation}
d_{L}(z)\equiv a_{0} \left(1+z \right) r(z)
\label{dL}
\end{equation}
and 
\begin{equation}
\Delta\left(m-M\right)(z)\equiv 
5 \log \left(\frac{d_{L1}(z)}{d_{L2}(z)} 
\right),
\label{DeltamM}
\end{equation}
where $d_{L1}$ is the luminosity distance in the spatially flat
VCDM model and
$d_{L2}$ is the luminosity distance in an arbitrary fiducial 
model
used as normalization. We will set $d_{L2}$ as the luminosity 
distance in an open
and empty Universe [$a(t)=t$ and $k=-1$], which 
is the convention used by
\citet{Riess01}.

\begin{figure}
\plotone{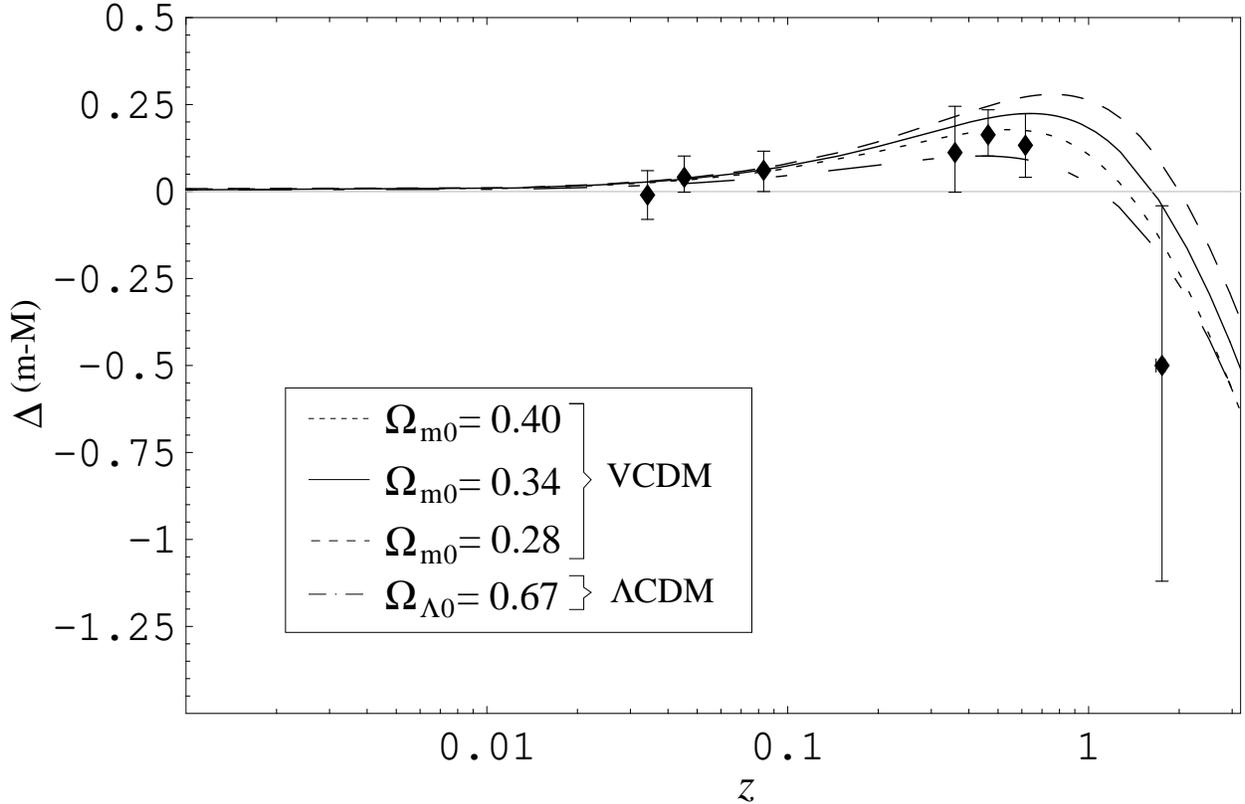}
\caption{Plot of the type-Ia supernovae distance modulus
(normalized to a spatially open and empty cosmos) as a
function of redshift $z$ for the spatially flat VCDM
(with $\Omega_{m0}=0.34\pm 0.06$) and $\Lambda$CDM (with
$\Omega_{\Lambda 0}=0.67$) models.}
\label{fig:SNe}
\end{figure}
It is important to note that the expression for
$\Delta\left(m-M \right)$ 
does not depend explicitly on the present value 
of the Hubble constant $H_0$. However, if we
adopt the results of the
previous section summarized in figure~\ref{fig:95cl}, then fixing 
different values of $H_0$ in that figure lead to different
$95\%$-confidence-level ranges for $\Omega_{m0}$, which in turn give
rise to different predictions for the SNe-Ia luminosity distances.
In particular, using for $H_0$ the best-fit value 
$H_0= 65$~${\rm km}~{\rm s}^{-1}~{\rm Mpc}^{-1}$ gives us 
$\Omega_{m0}=0.34 \pm 0.06$ (see fig.~\ref{fig:95cl}). 
In figure~\ref{fig:SNe} we plot the distance modulus as a function
of redshift predicted by the VCDM model using this
``best-fit range'' for $\Omega_{m0}$, as well as the observed 
distance moduli of SNe-Ia. It can be seen from figure~\ref{fig:SNe}
that fixing 
the values of $\Omega_{m0}$ and
$H_0$ that best fit the CMBR data also gives a very good 
no-parameter fit to the SNe-Ia data. 
Moreover, any value of 
$\Omega_{m0}$ in the ``best-fit range'' 
$\Omega_{m0}= 0.34 \pm 0.06$ gives a reasonably
good fit to the SNe-Ia data,
as shown by the dashed curves in figure~\ref{fig:SNe}.
We also show in figure~\ref{fig:SNe} the distance modulus predicted 
by the
$\Lambda$CDM model with $\Omega_{\Lambda 0}=0.67$. We see that
even though the predictions of both models differ significantly
in the range $0.5 \lesssim z \lesssim 1.5$, current data are still not
able to make a clear distinction between them.

More numerous and accurate data on SNe-Ia luminosity distance are
expected for the near future.
The planned Supernova Acceleration Probe
(SNAP)\footnote{http://snap.lbl.gov/}, for instance,
aims at cataloging up to 2,000 SNe-Ia per year
in the redshift range
$0.1\lesssim z \lesssim 1.7$. This improvement in our knowledge 
of the luminosity distances of SNe-Ia will provide a much 
stronger test of
the VCDM model.

\section{Number Counts}
\label{sec:GC}

Counting galaxies or clusters of galaxies as a function of their
redshift seems to be a very promising way to test different
cosmological models~\citep{Huterer01,Podariu01}.  
The idea behind this procedure is that once we know, either by
analytic calculations or by numerical simulations,
the evolution of the 
comoving (i.e., 
coordinate) density
of a given class of objects (e.g., galaxies or clusters of 
galaxies),
counting the observed 
number of such objects, 
per unit solid angle as a function of 
their redshift, is equivalent to tracing back the 
area of the Universe at different stages that we can observe today.
In other words, it is equivalent to determining our past light cone
by constructing it from the area of
these observed spherical sections of the Universe, parametrized
by their redshift.
Since this light cone is very sensitive to the 
underlying cosmological model,
number counts provide a valuable tool for testing the mechanism
which accounts for the accelerated expansion of the Universe.

This kind of test was first
performed using galaxies brighter than certain 
(apparent) magnitudes by \citet{Loh86},
with the simplified assumptions that
the comoving density of galaxies is
constant  and
that their luminosity function retains similar shape over the
redshift
range $0.15\lesssim z \lesssim 0.85$. Using the photometric redshift 
of
$406$ galaxies in that range, they were able to measure the 
ratio of the total energy density in the Universe to the
critical density, obtaining $\Omega_{0} = 0.9^{+0.6}_{-0.5}$.
However, the
validity of Loh \& Spillar's assumptions 
is still not clear due to the lack, 
to the
present, of a complete theory of galaxy formation and evolution.
In order to circumvent this problem,
\citet{Newman00} then suggested that galaxies having
the same circular velocity
may be regarded as good candidates for number
count tests, since the evolution of the
comoving number density of dark halos having a given
circular velocity
can be calculated by a semi-analytic approach. Moreover,
they claim, the comoving abundance of such objects at 
redshift $z\approx 1$ (relative to their
present abundance) is very insensitive to the underlying
cosmological model (under reasonable matter power spectrum 
assumptions). 
Other objects that one can count
are clusters of
galaxies~\citep{Bahcall98,Blanchard98,Viana99,Haiman01,Newman02}, 
which are simpler objects than
galaxies, in the sense that their formation and evolution, and 
therefore their
density, depend 
mostly on well-understood gravitational physics.

Whatever class of objects one uses to perform the number count
test, a key ingredient one needs to provide as an input, as 
stressed above,
is the evolution of their comoving
density,
\begin{equation}
n_c(z)\equiv \frac{\sqrt{1-kr^2}}{r^2}\frac{dN(z)}{dr d\Omega}\;,
\label{nc}
\end{equation} 
where $d\Omega \equiv  \sin \theta d\theta d\phi$ is the solid-angle
element and $dN(z)$ is the number of such objects,
at the spatial section at redshift $z$,
contained in the coordinate
volume $r^2dr d\Omega/\sqrt{1-kr^2}$.  In
order to find out the number of objects, per unit solid angle, with
redshift between $z$ and $z+dz$, we have to use the fact that the
objects we are observing today 
with redshift $z$ possess
coordinate $r$ which satisfies the past light-cone equations
(\ref{drdz}) and (\ref{rz}).
Thus, by making use of these equations to
eliminate the explicit radial dependence in equation~(\ref{nc}), we 
get the number of observed objects with redshift between $z$ and
$z+dz$, per unit solid angle:
\begin{equation} 
\frac{dN}{dzd\Omega}(z)dz=
\cases{
n_c(z)\left(a_0 H_0\right)^{-3}
E(z)^{-1}
\left(
\int_0^z E(z')^{-1}dz'
\right)^2 dz &, $k=0$ \cr
n_c(z)\left(a_0 H_0\right)^{-3}
E(z)^{-1}
\left[
\Omega_{k0}^{-1/2}
\sinh \left(
\Omega_{k0}^{1/2}\int_0^z E(z')^{-1}dz'
\right)
\right]^2 dz &, $k\neq 0$},
\label{dndz}
\end{equation}
with $E(z) \equiv H(z)/H_0$.
\begin{figure}
\plotone{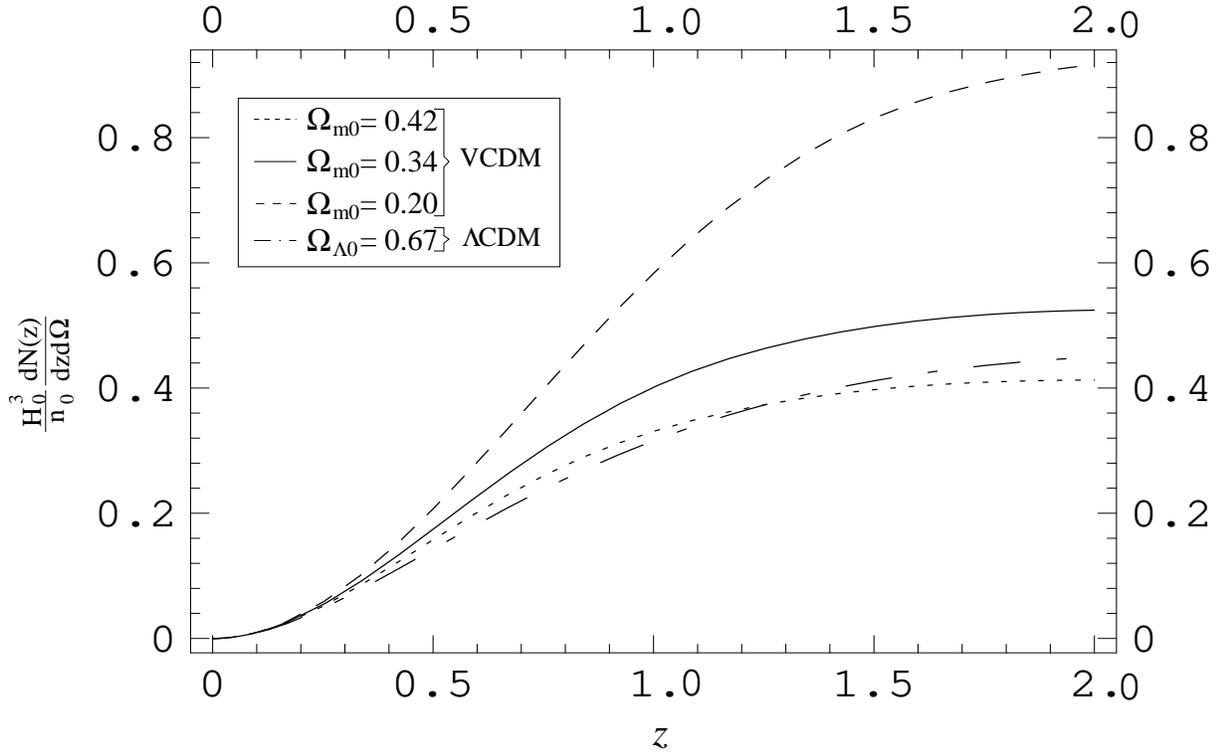}
\caption{Plot of the predicted number counts of objects per 
redshift interval and per solid angle, $dN(z)/dzd\Omega$ (normalized 
by the present number of such objects in the volume $H_0^{-3}$) as
functions of their redshift $z$, for the 
spatially flat VCDM (with 
$\Omega_{m0}=0.34^{+0.08}_{-0.14}$) and
$\Lambda$CDM (with $\Omega_{\Lambda 0}=0.67$) models. 
In this plot, the comoving density is assumed to be 
constant.}
\label{fig:dndz}
\end{figure}

In order to illustrate number counts predicted by the VCDM
cosmological model, in figure~\ref{fig:dndz} we plot 
$dN(z)/dzd\Omega$
given by
equation~(\ref{dndz}), 
with $H(z)$ given by equation~(\ref{Hz}). Then, we
apply the resulting formula to the spatially flat case 
and 
use the
parameters obtained in 
section~\ref{sec:CMBR} adopting the HST-Key-Project 
constraint 
(see fig.~\ref{fig:95cl}), namely $\Omega_{m0}=0.34^{+0.08}_{-0.14}$. 
Also, we follow \citet{Podariu01} and \citet{Loh86} in
assuming, for simplicity, the constancy of the comoving density,
$n_c(z)=n_0a_0^3$ ($n_0$ is the {\it proper} density at the present
epoch). For sake of comparison,
we also plot the spatially flat $\Lambda$CDM prediction, with 
$\Omega_{\Lambda 0}=0.67$ and 
the same assumption of 
constant density.
Obviously, the predictions could be improved by dropping
this latter assumption and taking into account
the density evolution of the observed objects, as mentioned earlier.
However, calculating such evolution is beyond the scope of the 
present paper, not to mention the fact that it is still not completely
clear which class of objects we should choose.
Moreover, once 
a more precise $n_c(z)$ is known, it is a simple task to take it into
account since $dN(z)/dzd\Omega$ is simply proportional to
$n_c(z)$.
From figure~\ref{fig:dndz} we see 
that the VCDM model predicts
more objects to be observed over redshifts 
$z\lesssim 2$ than the
$\Lambda$CDM model. In fact, in a small redshift interval
$\Delta z \ll 1$ around $z\approx 1$ the VCDM model predicts
approximately $30\%$ more objects than
the $\Lambda$CDM model for approximately the same value of
$\Omega_{m0}$ (and
the same value of $n_0/H_0^3$).
Note that this last conclusion should also hold for
the counts of galaxies at fixed circular velocities
suggested by \citet{Newman00}, 
since their comoving density,
even though not constant, is very insensitive to the underlying
cosmological model at $z\approx 1$. 
We did not mention
here the presence of selection effects,  
since they highly depend on the measurement
procedure itself. Notwithstanding, these effects may also be 
included in the computation via an ``effective'' 
$n_c(z)$, which then should be
viewed as the number of objects at the spatial section with
redshift $z$, per comoving volume, satisfying
the detectability conditions.

Measurements of $dN/dzd\Omega$ will provide a valuable way to 
test the VCDM cosmological model and distinguish it from the 
$\Lambda$CDM model, when combined with CMBR anisotropy results. 
Such measurements will soon become available, as the DEEP (Deep 
Extragalactic Evolutionary Probe) Redshift 
Survey\footnote{http://deep.ucolick.org/}
expects to complete
its measurements of the spectra of approximately $65,000$ 
galaxies 
in the redshift range
$0.7\lesssim z\lesssim 1.5$
by the year 2004.

\section{Vacuum Equation of State}
\label{sec:VES}

The dark-energy equation of state $\rho_v=\rho_v(p_v)$
in the VCDM cosmological model with $k=\pm 1$ or $0$
can be easely obtained, for $t > t_j$,
from equations~(\ref{rhov}) and (\ref{pv}):
\begin{equation}
\rho_v=3p_v+\frac{{\bar m}^2}{8\pi G}
\left[
1-
\left(
1+\frac{32 \pi G}{{\bar m}^2}p_v
\right)^{3/4}
\right]\;.
\label{rhop}
\end{equation}
Moreover, from the same pair of equations we also obtain the
ratio $w\equiv p_v/\rho_v$ as a function of redshift:
\begin{eqnarray}
w(z)&=&\frac{\zeta^4-1}{3\zeta^4-4\zeta^3+1}\nonumber \\
&=&\frac{\zeta^3+\zeta^2+\zeta+1}{3\zeta^3-\zeta^2-\zeta-1}\;,
\;\;0<\zeta<1\;,
\label{wz}
\end{eqnarray}
where $\zeta \equiv a_j/a=(1+z)/(1+z_j)$.
Note that equations~(\ref{rhov}), (\ref{pv}),
(\ref{rhop}), and (\ref{wz})
are the same as the respective 
ones presented by
\citet{Parker01} in dealing with the spatially flat VCDM 
model.
(Note, however, that the spatial curvature changes the value
of $z_j$; see eqs.~[\ref{zj}] and [\ref{mHdimensionless}].)
\begin{figure}
\plotone{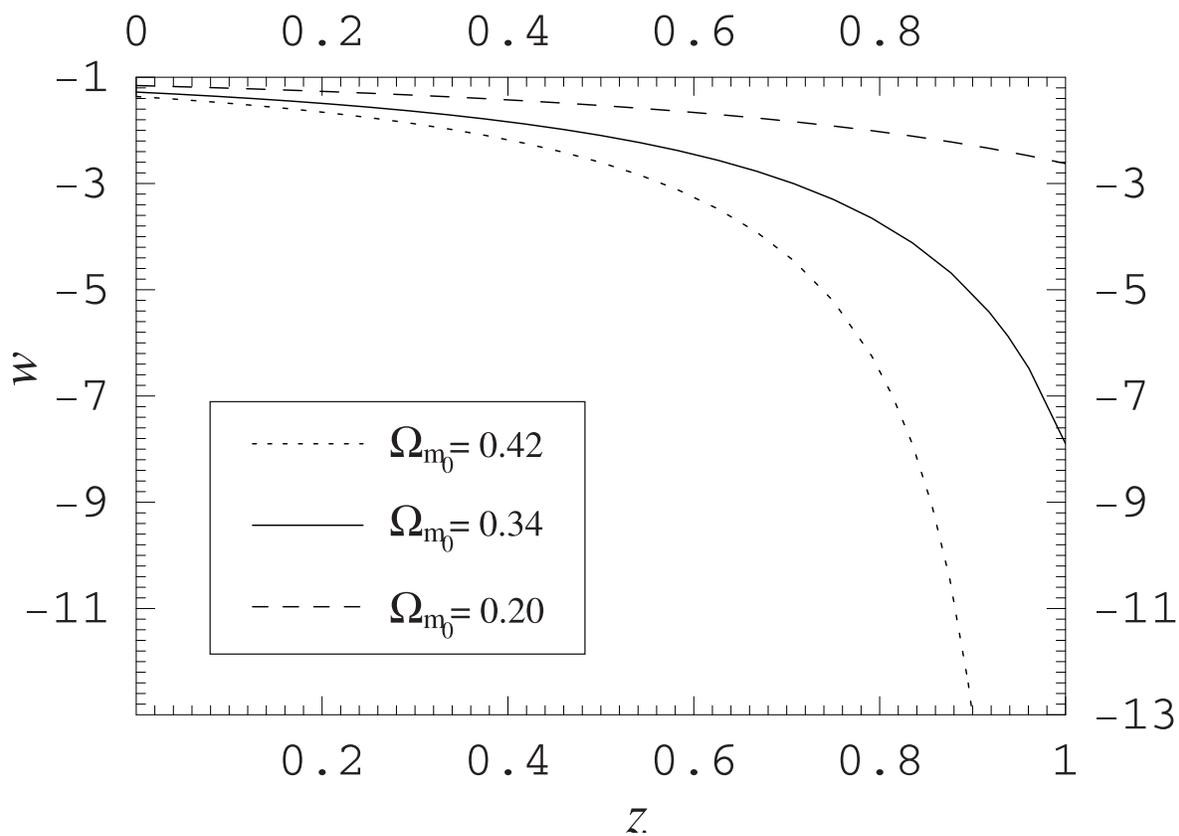}
\caption{Plot of the predicted ratio $w\equiv p_v/\rho_v$ as a
function of redshift $z$, for the 
spatially flat VCDM model with 
$\Omega_{m0}=0.34^{+0.08}_{-0.14}$
(the HST-Key-Project constraint was adopted). 
The present value of such
ratio is $w_0 = -1.28^{-0.08}_{+0.12}$. Moreover,
$w\to -\infty$ as $z\to z_{j-}=(1.19^{-0.19}_{+0.47})_-$}
\label{fig:wz}
\end{figure}

\begin{figure}
\plotone{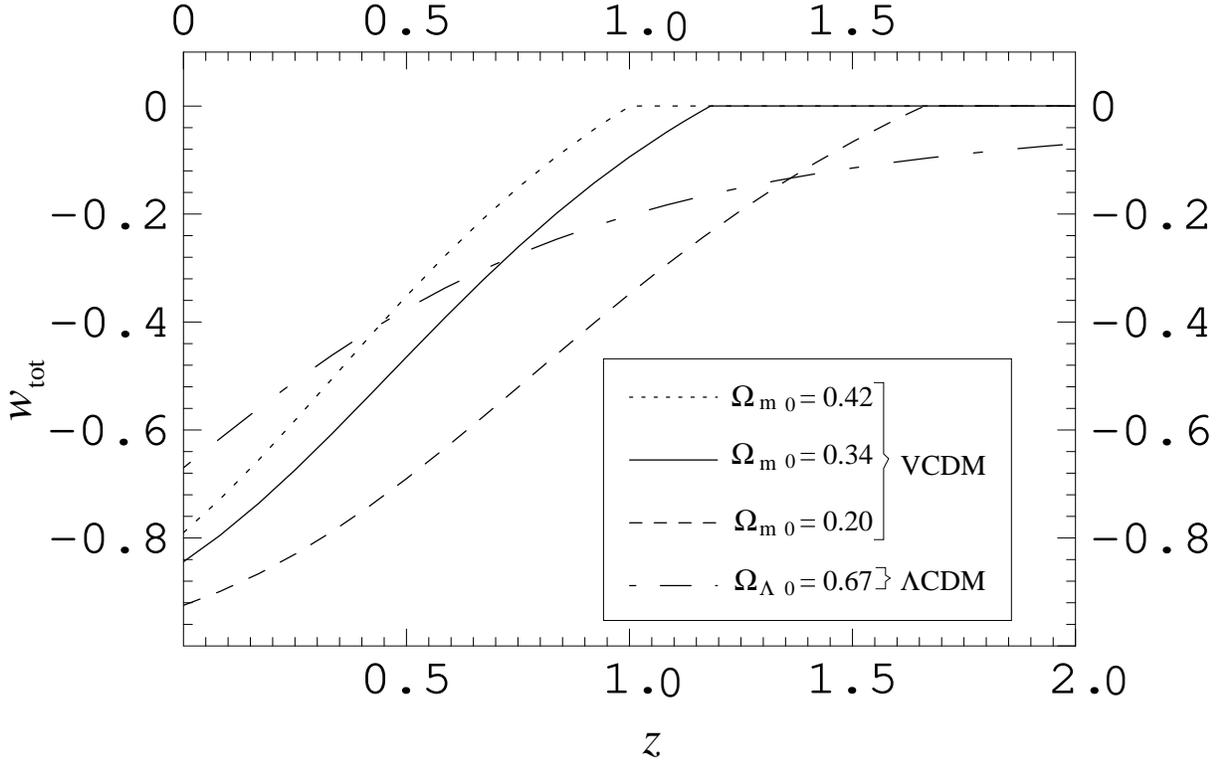}
\caption{Plot of the ratio $w_{\rm tot}\equiv p/\rho$ as a
function of redshift $z$, for the 
spatially flat VCDM model with 
$\Omega_{m0}=0.34^{+0.08}_{-0.14}$ and
for the spatially flat 
$\Lambda$CDM model with $\Omega_{\Lambda 0}=0.67$.
With these parameters, the 
present values of $w_{\rm tot}$ and  $w^{(\Lambda)}_{\rm tot}$
are  $w_{\rm tot}(z=0)=-0.85^{+0.06}_{-0.07}$ and
$w^{(\Lambda)}_{\rm tot}(z=0)=-0.67$.
Note that for the VCDM model the ratio $w_{\rm tot}$
is negligible for times earlier than $t_j$ (i.e., $z>z_j$), during 
the matter dominated era.}
\label{fig:wT}
\end{figure}

In figure~\ref{fig:wz} we plot the redshift dependence of the ratio 
$w$,
given by equation~(\ref{wz}),
for the spatially flat VCDM model
using the cosmological
parameters obtained in section~\ref{sec:CMBR} adopting
the HST-Key-Project constraint.
The present value of this ratio,
$w_0\equiv w(z=0)$, using $\Omega_{m0}=
0.34^{+(0.46;0.08)}_{-0.14}$, is 
$w_0=-1.28^{-(0.91;0.08)}_{+0.12}$. 
Note, from the expression for $w(z)$, that 
$w\to -\infty$ as $z\to z_{j-}$, 
which is
simply a consequence of the previously mentioned fact that
the vacuum energy approaches a
negligibly small value
(more rapidly than the vacuum pressure) as $z \to z_{j-}$.
That no drastic consequence follows from the divergence of $w$ 
is evident from figure~\ref{fig:wT}, where we plot,
using equations (\ref{prhoeqofstate}) and (\ref{rhoa4}),
the ratio
$w_{\rm tot} \equiv  p/\rho$ between the {\it total} pressure $p$ 
and the 
{\it total} energy density $\rho$ present in the Universe, as a 
function of redshift. 
For the sake of comparison,
we also plot in the same figure the correspondent 
ratio $w^{(\Lambda)}_{\rm tot}$
given by the spatially flat
$\Lambda$CDM model,
\begin{eqnarray}
w^{(\Lambda)}_{\rm tot}&=&\frac{p_r+p_\Lambda}{\rho_{r}+
\rho_m+\rho_\Lambda} 
\nonumber \\
&=& \frac{(\Omega_{r0}/3)(1+z)^4-\Omega_{\Lambda 0}}
{\Omega_{r0}(1+z)^4 +(1-\Omega_{r0}-\Omega_{\Lambda 0})
(1+z)^3+\Omega_{\Lambda 0}}\;,
\label{wTlambda}
\end{eqnarray}
with $\Omega_{\Lambda 0}=0.67$ and $\Omega_{r0}=8.33 \times 10^{-5}$.
In the VCDM model note that for times earlier than $t_j$ (i.e., 
$z> z_j$), during
the matter dominated era,
the total energy density, $\rho\approx 
\rho_{m}$, and the total pressure, $p=
\rho_{r}/3$, lead to a negligible value of the ratio $w_{\rm tot}$.
After $t_j$ ($z<z_j$), the negative pressure of the 
vacuum grows very rapidly in magnitude, becoming dominant
and determining a very sharp transition to the
dark-energy dominated era. This is an important distinction between 
the 
VCDM (i.e., vacuum metamorphosis) model 
and the $\Lambda$CDM model, which presents a rather gradual transition
(see fig.~\ref{fig:wT}). 
\begin{figure}
\plotone{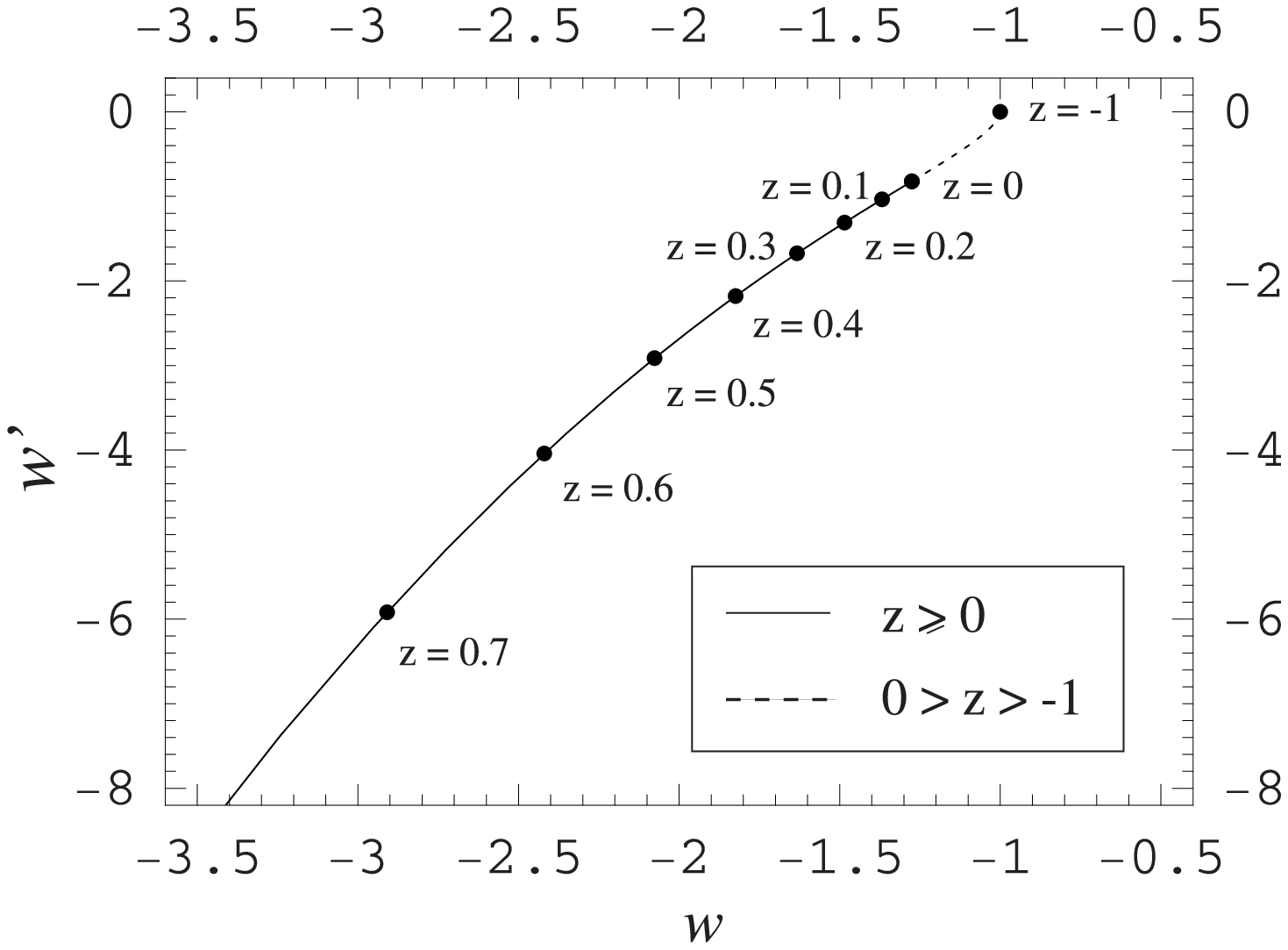}
\caption{Plot of the curve $\left(w(z),w'(z) \right)$,
parametrized by the redshift $z$,
for the 
spatially flat VCDM model with 
$\Omega_{m0}=0.34$. The solid-line curve 
corresponds to redshift $z\geq 0$ while the dashed-line one
corresponds to $-1<z<0$. At the present epoch we have
$\left(w_0,w'_0 \right)=\left(-1.28,
-0.8\right)$. Note that 
$\left(w(z),w'(z) \right)\to \left(-1,0 \right)$ as
$z\to -1$, which means that the dark energy of the VCDM model
behaves like a cosmological constant in the asymptotic future.}
\label{fig:dwdz}
\end{figure}

In order to analyze not only the value of $w$ but also its rate of
change in redshift, we plot in figure~\ref{fig:dwdz},
using $\Omega_{m0}=0.34$,
the
curve $\left(w(z),w'(z) \right)$, where
$w'(z)\equiv dw(z)/dz$. The redshift $z$ is used as the 
parameter of the curve. The present value $w'_0\equiv 
w'(z=0)$ 
predicted by the spatially flat VCDM model is
$w'_0=-0.8^{-(4.2;0.3)}_{+0.4}$. Note that $w \to -1$ and 
$w'\to 0$ as $z\to -1$, which means that
in the asymptotic future (assuming that 
nothing new will prevent the unbounded expansion of the Universe)
the dark energy of the VCDM model behaves like an effective
cosmological constant, with value given by 
$\Lambda_{\rm eff} \equiv {\bar m}^2/4=
5.1^{-(3.2;0.4)}_{+2.1}\times 10^{-66}~{\rm eV}^2$.
For comparison,
in a $\Lambda$CDM model with
$H_0\approx 65$~${\rm km}~{\rm s}^{-1}~{\rm Mpc}^{-1}$ and 
$\Omega_{\Lambda 0}\approx 0.67$, one would have 
$\Lambda = 3 
\Omega_{\Lambda 0} H_0^2\approx 3.8 \times 10^{-66}~{\rm eV}^2$.

The experimental determination of $w(z)$, avoiding model-dependent
assumptions, 
relies basically on measurements that, at least in principle, will
determine $H(z)$ with sufficient precision to
provide also a reliable determination of $ H'(z)\equiv
dH(z)/dz$~\citep{Huterer01}. To see this, let us consider the 
conservation equation satisfied by the total energy density 
$\rho$ and the total pressure $p$, namely
$d(\rho a^3)+p\; d
\left(a^3\right)=0$. Thus, with the (only)
assumption that matter and radiation are separately conserved, we
have that the energy density, $\rho_X$, and pressure, $p_X$, 
of dark energy (whatever it is) also satisfy the same conservation 
equation, which implies
\begin{eqnarray}
\frac{1}{\rho_X}\frac{d\rho_X}{dz}=
-\frac{(1+w_X)}{a^3}\frac{d\left(a^3\right)}{dz}
= 3\frac{(1+w_X)}{(1+z)}\;,
\label{drho}
\end{eqnarray}
where $w_X\equiv p_X/\rho_X$
and we have used $1+z=a_0/a$. Considering the general expression
for the Hubble parameter as a function of redshift
(which is obtained from
Einstein's equation together with the assumption of separate
conservation of matter and radiation),
\begin{equation}
\frac{H(z)^2}{H_0^2}=\Omega_{r0}(1+z)^4 +\Omega_{m0}(1+z)^3
+\Omega_{k0}(1+z)^2 +\Omega_{X0}\frac{\rho_X(z)}{\rho_{X0}}\;
\label{HzX}
\end{equation}
[with $\rho_{X0}$ being the present value of the dark-energy density
and $\Omega_{X0}\equiv 8\pi G\rho_{X0}/(3H_0^2)$],
and using it and
its redshift derivative 
to evaluate the left-hand-side of
equation~(\ref{drho}), we finally
obtain the desired expression for $w_X(z)$:
\begin{eqnarray}
w_X(z)&=&\frac{(1+z)}{3}\frac{\left[d\rho_X(z)/dz\right]}
{\rho_X(z)}-1
\nonumber \\
&=& 
\frac{1}{3}
\frac{\left[2(1+z) E'(z)-3E(z)\right] E(z)-
(1+z)^2\left[\Omega_{r0}(1+z)^2-\Omega_{k0}\right]}
{E(z)^2-\Omega_{r0}(1+z)^4-\Omega_{m0}(1+z)^3-
\Omega_{k0}(1+z)^2}\;,
\label{wXz}
\end{eqnarray}
where, again, $E(z)\equiv H(z)/H_0$ and $ E'(z)\equiv dE(z)/dz$.
Thus, as stated above, $w_X(z)$ can be found from the
determination of
$H(z)$ and $H'(z)$. 
The quantity $H(z)$ can be determined from the direct observables
$d_L(z)$ and
$dN(z)/dzd\Omega$, and the quantity $n_c(z)$
\citep{Huterer01}. 
This is done by using equation~(\ref{drdz}) to express the derivative
with respect to $r$ in equation~(\ref{nc}) in terms of a derivative
with respect to $z$, and then using equation~(\ref{dL}) to express 
$r(z)$ in terms of $d_L(z)$. This leads to the following expression
for $H(z)$:
\begin{equation}
H(z)=\frac{n_c(z)}{a_0^3}
\left(
\frac{dN(z)}{dzd\Omega}
\right)^{-1}
\frac{d_L(z)^2}{(1+z)^2}\;.
\label{HzdLnc}
\end{equation}
Thus, by 
considering measurements of luminosity distances and number counts,
$H(z)$ can be regarded as a directly observable quantity, which
gives $w_X(z)$ through equation~(\ref{wXz}).
In this sense, future data provided by the proposed 
satellite SNAP on supernovae 
luminosity distances (see sec.~\ref{sec:SNe})
and by the DEEP redshift 
survey on number counts (see sec.~\ref{sec:GC}) may
greatly improve our knowledge of the dark-energy equation of state.

\section{Conclusion}

We have shown that the current observational data
indicating
that the expansion of the Universe is undergoing 
acceleration are quite consistent with 
the hypothesis that a transition to
a constant-scalar-curvature 
stage of the expansion occurred at a redshift $z\sim 1$ in the 
spatially flat FRW universe having zero cosmological constant.
This is the scenario proposed in 
the VCDM (or vacuum metamorphosis)
model introduced by Parker and Raval. The late constancy of 
the scalar curvature at a value 
$R_j= \mbar^2$ is induced by quantum effects of a 
free scalar field of low mass
in the curved cosmological background. The
parameter $\mbar$, related to the mass of the field, is the only new
relevant 
parameter introduced in this model, 
and can be expressed in terms of the present
cosmological parameters $H_0$, $\Omega_{m0}$, $\Omega_{r0}$, and
$\Omega_{k0}$ (see eq.~[\ref{mHdimensionless}]).

Comparison of the CMBR-power-spectrum data 
with the flat-VCDM-model
prediction, without or with the HST-Key-Project result
as a constraint
(see figs.~\ref{fig:CMBR} and \ref{fig:95cl},
and table~\ref{tab:peaks}), gives the 
values of the cosmological parameters to be
$H_0=65^{-(16;1)}_{+10}$~${\rm km}$~${\rm s}^{-1}$~${\rm Mpc}^{-1}$ 
and
$\Omega_{m0}=0.34^{+(0.46;0.08)}_{-0.14}$. (Recall the definition of 
our notation in sec.~\ref{sec:mbar}:
the uncertainties appearing in parenthesis refer to the $95\%$ 
confidence level, without and with
the HST constraint, respectively.)
Such values lead to
$\mbar= 4.52^{-(1.76;0.18)}_{+0.84} 
\times 10^{-33}$~${\rm eV}$, and the best-fit values from
the CMBR data
give rise
to a very
good no-parameter fit to the SNe-Ia observational data.
However, the SNe-Ia data are not accurate enough
to draw a clear distinction between the VCDM and $\Lambda$CDM models.
Other quantities of interest predicted by the 
VCDM model with the cosmological parameters mentioned above
are 
the time and redshift at the transition between the 
matter-dominated
and constant-scalar-curvature stages,
($t_j=5.33^{+(3.41;0.22)}_{-0.84}$~${\rm Gyr}$ and
$z_j=1.19^{-(0.76;0.19)}_{+0.47}$),
the time and redshift when the accelerated expansion started
($t_a=8.0^{+(5.2;0.4)}_{-1.2}$~${\rm Gyr}$ and 
$z_a=0.67^{-(0.59;0.15)}_{+0.35}$),
and
the age of the Universe, 
$t_0=14.9\mp 0.8$~${\rm Gyr}$. 

Regarding future tests of the VCDM model, we have presented the 
prediction of number counts as a function of redshift, and compared it
with the analogous $\Lambda$CDM prediction (see fig.~\ref{fig:dndz}).
For approximately the same cosmological parameters, the VCDM model
predicts nearly $30\%$ more objects 
to be observed in a small redshift interval around $z\approx 1$ than
the $\Lambda$CDM model. Data provided by the DEEP Redshift Survey 
in the
near future will likely be able to distinguish these two models. 
Also, DEEP data combined with future measurements of SNe-Ia
luminosity distances provided by the proposed SNAP satellite
should greatly improve our knowledge of the dark energy equation of 
state, which bears the most distinct feature of the VCDM model:
$w<-1$ and $w'<0$ (see figs.~\ref{fig:wz} and \ref{fig:dwdz}). 

It should be noted that we have here considered the simplest 
form of the VCDM model, in which the transition to constant 
scalar curvature is continuous and effectively instantaneous
(see fig.~\ref{fig:wT}).
This form of the model makes definite predictions regarding the
distance moduli of SNe-Ia and number counts. Thus, it is
encouraging that it remains a viable model when confronted with
the current observational data. Other natural parameters that may 
come into the VCDM model are the time interval over which the 
transition occurs, and the vacuum expectation value of the scalar 
field. A nonzero value of the transition time interval would mainly 
affect the predictions around $z\sim 1$, 
and a nonzero value of the vacuum expectation value is likely to 
increase the ratio of pressure to density, $w$. Future observational 
data will determine if it is necessary to consider nonzero values 
for these parameters.

\acknowledgments

This work was supported by NSF grant PHY-0071044 and
Wisconsin Space Grant Consortium. The authors 
thank Koji Uryu for helpful 
comments and suggestions on the figures, and Alpan Raval
for helpful discussions.

\end{document}